\documentclass[a4paper,11pt]{article}
\usepackage{epsfig}
\usepackage{graphicx}
\usepackage{rotating}
\usepackage{amssymb}
\usepackage{amsmath}
\usepackage{subfigure}
\usepackage{multirow}
\def\beq{\begin{eqnarray}}
\def\eeq{\end{eqnarray}}
\def\be{\begin{equation}}
\def\ee{\end{equation}}
\usepackage[T1]{fontenc}
\usepackage[utf8]{inputenc}
\usepackage{authblk}
\newtheorem{thm}{Theorem}

\begin{document}

\title{Absence of first order transition in the random crystal field Blume-Capel model on a fully connected graph}

\author[*]{Sumedha;}
\author[**]{Nabin Kumar Jana}
\affil[*]{School of Physical Sciences, National Institute of Science Education and Research(NISER), Khordha, Jatni, Odisha-752050, India}

\affil[**]{ School of Mathematical Sciences, National Institute of Science Education and Research(NISER), Khordha, Jatni, Odisha-752050, India}





\date{\today}
\maketitle




\begin{abstract}
In this paper we solve the Blume-Capel model on a complete graph in the presence of random crystal field with a distribution, $P(\Delta_i) =p\delta(\Delta_i-\Delta)+(1-p) \delta(\Delta_i+\Delta)$, using large deviation techniques. We find that the first order transition of the pure system is destroyed for $0.046<p<0.954$ for all values of the crystal field, $\Delta$. The system has a line of continuous transition for this range of $p$ from $-\infty <\Delta <\infty$. For values of $p$ outside this interval, the phase diagram of the system is similar to the pure model, with a tricritical point separating the line of first order and continuous transitions. We find that in this regime, the order vanishes for large $\Delta$ for $p<0.046$(and for large $-\Delta$ for $p>0.954$) even at zero temperature.
\end{abstract}

\section{Introduction}

Presence of quenched randomness can drastically change the properties of a system. For systems that undergo continuous transition in their pure state, Harris \cite{harris} showed that the necessary condition for the disorder to be irrelevant is that the specific heat critical exponent $\alpha$, should be less than 
zero. Extending Harris's arguments, Imry and Wortis\cite{imrywortis} and Aizenmann and Wehr\cite{aizenman} showed that the Gibbs state is always unique in the presence of disorder in two dimensions. Hence, an infinitesimal disorder should either destroy the order completely or change it to continuous transition. This was further supported via real space renormalization group calculation in two dimensions\cite{huiberker,branco}. Recently, the result has been made rigorous for a large number of classical and quantum systems in two dimensions \cite{greenblatt}.  For $d>2$, in \cite{cardy}, via mapping of the interface between the ordered and disordered state to the random field Ising model, it was conjectured  that there is an impurity threshold below which the transition will retain the first order character. 

While the picture for $d \le 2$ is now pretty clear, we do not have a clear picture in the higher dimensions. Hence, understanding quenched disorder for $d>2$ in systems with first order transition in the pure state, is an important open problem. In this paper we study a spin-1 model, which has a continuous transition in the absence of crystal field, but has a rich phase diagram with regions of first and second order transitions, separated by a tricritical point, in the presence of crystal field \cite{blume, capel,beg}.

Spin-1 model with crystal-field, also known as Blume-Capel model, was introduced by Blume\cite{blume} and Capel\cite{capel} separately to explain the first order magnetic transition in materials like $UO_2$\cite{frazer}. The model has been very successful in explaining many interesting physical phenomena. For example, critical behaviour of $He_3-He_4$ mixture in random media is modelled well by Blume-Capel model with random crystal field\cite{helium,buzano}. Recently, Blume-Capel model has been used to study the  phenomena of inverse melting\cite{schupper}. Also, shear induced  rigidity in granular materials has been studied via a mapping to the Blume-Capel model\cite{bulbul}.

Blume-Capel model has a rich phase diagram \cite{beg} and it has been one of the most well studied models in statistical mechanics. Effect of random crystal field has also been studied using various approximation techniques like mean field theory\cite{mfdilution}, effective field theory\cite{yuksel}, Bethe lattice\cite{bethe}, pair approximation\cite{lara}, renormalization group\cite{branco}, hierarchical lattices\cite{snowman} and replica method\cite{santos}. While for $d=2$ using real space renormalization group\cite{branco}, it was shown that the system  has only a continuous transition in the presence of random field, in agreement with the known rigorous results, situation in higher dimension is not that clear. Different methods do not agree with each other in their prediction of the phase diagram. For example, replica method\cite{santos} predicts a ferromagnetic state at zero temperature for all strengths of disorder. This is in contrast to predictions for the same model using pair approximations \cite{lara}. Also it has been shown that the predictions of effective field theory \cite{kaneyoshi} and mean-field theory\cite{mfdilution} do not match at low temperatures. Simulations are challenging even for the pure case\cite{zierenberg}, and with disorder averaging it is a formidable task\cite{fernandez,fytas}. Hence, exact solutions are important.

Here we present an exact solution of the Blume-Capel model with random crystal field on a complete graph using the method of large deviations\cite{touchette}. Pure version of the Blume-Capel model on a fully connected graph was studied earlier using large deviation techniques\cite{ellis, barre}. The method used,  though cannot be extended to the case with disorder. Our method relies on recognising that the disorder average can be done on the non interacting part for a given magnetisation, as the interacting part of the Hamiltonian depends only on the total magnetisation. We have recently used it to study the spin $1/2$, $p$-spin interaction  model\cite{sumedha}. In this paper, we extend the method to the Blume-Capel model. We reproduce the earlier known results for pure Blume-Capel model. In the presence of  random crystal field, we find that beyond a threshold, the first order transition disappears completely (in agreement with renormalization group flow predictions\cite{cardy}). For weak disorder($p<0.046$ and $p>0.954$), the phase diagram is similar to the pure Blume Capel model. In constrast to earlier studies \cite{mfdilution,santos}, we find that the line of first order transitions end at a finite $\Delta$ at zero temperature.

Plan of the paper is as follows: In Section 2 we define the model and give the general method of solution in Section 3. In Section 4 , we solve the pure Blume-Capel model using large deviations and in Section 5 we solve it with bimodal random crystal field. We conclude in Section 6.

\section{Model}

We consider the Blume-Capel model on a complete graph, with the Hamiltonian given by:
\begin{equation}
\label{hamiltonian}
H(C_N) = -\frac{1}{ 2N} (\sum_{i} s_i)^2 -\sum_{i} \Delta_{i} s_{i}^2
\end{equation}
where $C_N=(s_1,s_2,\ldots,s_N)$ denotes a configuration of spin variables $s_i$. Each spin $s_i$ takes  three values, $\pm1,0$, and $\Delta_i$ are the external random crystal fields. A positive $\Delta_i$ will favour a $\pm 1$ spin at a site, while a negative $\Delta_i$ will favour a $0$ spin at a site. Hence, in the absence of disorder, the system will approach Ising model for $\Delta \rightarrow \infty$. This system has two order parameters: magnetisation, $m=<s>$ and  magnetic particle density(or quadrulpolar moment), $q=<s^2>$. The probability of a particular configuration $C_N$, for a realization of random field $\{\Delta_i\}$ 
would be given by

\begin{equation}
P_{N,\beta}(C_N,\{\Delta_i\})=\frac{\exp \left(\frac{\beta}{2 N} (\sum_i s_i)^2+\beta \sum_{i} \Delta_{i} s_{i}^2\right)}{Z_{N,\beta}}
\end{equation}
where $\beta$ is the inverse of temperature. And $Z_{N,\beta}$ is the normalisation, also known as  the partition function, for a given realization of $\{ \Delta_i \}$. 
Here the sequence of both the order parameters as a whole for $N$ particle system satisfies large deviation principle(LDP) with respect to the sequence of probabilities $\{P_{N,\beta}\}$ and this result holds, almost surely with respect to the realisation of the random crystal fields under certain assumptions in $\{\Delta_i\}$. Moreover, we will show in Section \ref{method}, that the rate function for the LDP can be calculated through tilting the Gartner-Ellis theorem\cite{hollander}.

Note that if a sequence of random variables $\{Y_N\}$ taking values in a complete separable metric space $\mathcal{X}$ satisfies LDP with respect to a sequence of probabilities 
$\{\mu_N\}$ on $\mathcal{X}$ with rate $N$ and continuous rate function  $R:\mathcal{X}\rightarrow \mathbb{R}$, then for any Borel set $A$ of $\mathcal{X}$,

\begin{equation}
	\mu_N(Y_N\in A)\sim e^{-NR(A)},
\end{equation}
where $R(A)=\inf\{R(z):\, z\in A\}$. Since $\mu_n$'s are probabilities, $R(z)\geq 0, \quad \forall z\in \mathcal{X}$ and $Y_N$'s will get more likely values where $R$ takes minimum values. We use this information to extract the behaviour of the physical observables of the system by extending the method of \cite{sumedha,lowe} to this model.

\section{Method}
\label{method}

We first give the general results for the model and in Section 3 and 4 we will solve it for a given  $\{\Delta_i\}$.

Let us first consider the non interacting part of the Hamiltonian, $H_{ni}(C_N)=-\sum_{i} \Delta_i s_i^2$. Also, let $Q$ be the product measure on $\{-1,0,1\}^\mathbb N$ generated by $Q_i$, where $Q_i(\{1\})=Q_i(\{-1\})\propto \exp(\beta\Delta_i)$ and $Q_i(\{0\})\propto 1$ for $i\in \mathbb N$ and $\beta>0$.  The logarithmic moment generating function of  $N Y_N=(N S_N,N \sigma_N)$, where $S_N=\sum_i s_i/N$ and $\sigma_N= \sum_i s_i^2/N$ with respect to $Q$ is then
\begin{eqnarray}
\Lambda_N (x_1,x_2) &=& \log E_Q\exp\{N(x_1S_N+x_2\sigma_N)\}\\ \nonumber 
&=&\sum_{i=1}^N\log E_{Q_i}\exp\{(x_1s_i+x_2s_i^2)\}.
\end{eqnarray}
Then
\begin{equation}
\frac{1}{N}\Lambda_N(x_1,x_2)=f_N(x_1,x_2)-f_N(0,0),
\end{equation}
where 
\begin{eqnarray}
f_N(x_1,x_2) &=& \frac{1}{N}\sum_{i=1}^N\log[2\exp(x_2+\beta\Delta_i) \cosh x_1+1].
\end{eqnarray}

Let us assume that $f_N\rightarrow f$ as $N\rightarrow\infty$ almost surely on $\mathbb{R}^2$ to some extended real valued function $f$ so that:

A1. $(0,0)$ is in the interior of the set $D_f=\{(x_1,x_2)\in \mathbb{R}^2:\quad f(x_1,x_2)<\infty\}$, 

A2. $f$ is a lower semi-continuous and is differentiable on interior of $D_f$,

A3. for any boundary point $(x_1,x_2)$ of $D_f$, $\lim_{(y_1,y_2)\rightarrow (x_1,x_2)}|\nabla f(y_1,y_2)|=\infty$.

Gartner-Ellis theorem \cite{touchette} implies that $\{N Y_N\}$ almost surely satisfies large deviation principle(LDP) with respect to $Q$ and the probability that $S_N=x_1$ and $\sigma_N=x_2$ in the large $N$ limit is given by the associated rate function:

\begin{equation}
R(x_1,x_2)=\begin{cases}\displaystyle\sup_{(y_1,y_2)\in\mathbb{R}^2}\{ x_1y_1+x_2y_2-\Lambda(y_1,y_2)\},& |x_1|\leq x_2, 0\leq x_2\leq 1\\ \infty, & \mbox{ elsewhere}\end{cases}
\end{equation}
where 
\begin{equation}
\Lambda(x_1,x_2)=f(x_1,x_2)-f(0,0).
\end{equation}

For a given distribution of the random crystal fields $\{ \Delta_i \}$, we can use the strong law of large numbers in the large N limit. The function $f(x_1,x_2)$ then is just the expectation value of $\log[2 \exp(x_2+\beta \Delta_i) \cosh x_1+1]$ with respect to the given distribution of crystal fields.

Now let us consider the probability of the configuration $C_N$ when the full Hamiltonian (Eq. 1) is considered:
\begin{equation}
P_{N,\beta}(C_N)\propto \exp(-\beta H(C_N)). 
\end{equation}

Note that $Y_N$ satisfies LDP with respect to $P_{N,\beta}$ also. Since induced probabilities of $P_{N,\beta}$ on $\mathbb{R}^2$ by $Y_N$ can be obtained by tilting the induced probabilities of $Q$ on $\mathbb{R}^2$ by $Y_N$, with the following bounded continuous function $F:\mathbb{R}^2 \rightarrow \mathbb{R}$:
\begin{equation}
F(x_1,x_2)=\begin{cases}
\frac{1}{2}\beta x_1^2, & 0\leq |x_1| \leq x_2 \leq 1,\\
\frac{1}{2}\beta x_2^2, & x_2 \leq 1 \& |x_1|>x_2,\\
\frac{1}{2}\beta\min\{1, x_1^2\}, & x_2 > 1.\\
\end{cases}
\end{equation}
Hence $F$ is just a bounded continuous extension of $\beta x_1^2/2$ on $0 \le |x_1| \le x_2 \le 1$, which is needed to apply the tilted large deviation principle \cite{hollander}. By Tilted LDP  we find that, $Y_N$ satisfies LDP with respect to $P_{N,\beta}$ with the rate function 
\begin{equation}
I(x_1,x_2)=R(x_1,x_2)-F(x_1,x_2)-\inf_{(y_1,y_2)\in \mathbb{R}^2}\{R(y_1,y_2)-F(y_1,y_2)\},
\label{fullratefunction}
\end{equation}
for $0\leq |x_1| \leq x_2 \leq 1$ and $\infty$ elsewhere. This implies that $P_{N,\beta}(\{C_N:Y_N(C_N)=(x_1,x_2)\}) \sim \exp(-N I(x_1,x_2))$. For the sake of simplicity of presentation, we have moved the derivation of Eq. \ref{fullratefunction} to the Appendix. Interestingly, while $R(x_1,x_2)$ is strictly a convex function, $I(x_1,x_2)$ need not be convex. 

As discussed in \cite{sumedha}, the rate function $I(x_1,x_2)$ in Eq. \ref{fullratefunction} is like the Landau free energy functional, whose minima gives the actual free energy of the system. Hence, we can obtain the phase diagram of the system by looking for the minima of $I(x_1,x_2)$ with respect to $x_1$ and $x_2$. This is what we will do in rest of the paper.

\section{Pure case}
Let us first calculate for the case of no disorder. The Hamiltonian of the system is 
\begin{equation}
H = -\frac{1}{ 2N} (\sum_{i} s_i)^2 -\Delta \sum_{i} s_{i}^2
\label{pure}
\end{equation}
In this case we get:
\begin{equation}
\Lambda(y_1,y_2)=\log[e^{\beta \Delta+y_1+y_2}+e^{\beta \Delta+y_2-y_1}+1]-\log[1+2e^{\beta \Delta}]
\end{equation}
Hence, to get $R(x_1,x_2)$ we need to find the supremum of 
\begin{eqnarray}
W(y_1,y_2)&=&x_1y_1+x_2y_2-\log[e^{\beta \Delta+y_1+y_2}+e^{\beta \Delta+y_2-y_1}+1]\\ \nonumber
&+&\log[1+2e^{\beta \Delta}]
\end{eqnarray}

On minimising $W(y_1,y_2)$ over all possible values, we get the equations for the supremum 
point $(y^*_1,y^*_2)$ to be

\begin{equation}
x_1=\frac{e^{\beta \Delta+y^*_2+y^*_1}-e^{\beta \Delta +y^*_2-y^*_1}}{1+e^{\beta \Delta+y^*_2+y^*_1}+e^{\beta \Delta +y^*_2-y^*_1}}
\end{equation}
and
\begin{equation}
x_2=\frac{e^{\beta \Delta+y^*_2+y^*_1}+e^{\beta \Delta+ y^*_2-y^*_1}}{1+e^{\beta \Delta+y^*_2+y^*_1}+e^{\beta \Delta +y^*_2-y^*_1}}
\end{equation}

Solving them simultaneously results in a particular simple equation for the supremum, $\{ y^*_1,y^*_2\}$.
\begin{eqnarray}
y^*_1 &=& \tanh^{-1} \frac{x_1}{x_2}\\
y^*_2 &=& -\beta \Delta+\log[(x_2 \mbox{sech} y^*_1)/(2(1-x_2))]
\end{eqnarray}

The $R(x_1,x_2)$, then comes out to be:
\begin{equation}
\label{entropy}
R(x_1,x_2)=x_1 y^*_1+x_2 y^*_2-\log[1/(1-x_2)]+\log[1+2 e^{\beta \Delta}]
\end{equation}

Hence $I(x_1,x_2)$, the rate function corresponding to the full Hamiltonian, Eq. \ref{pure}, would be given by Eq. \ref{fullratefunction}, i.e, 
\begin{equation}
I(x_1,x_2)=R(x_1,x_2)-\frac{\beta x_1^2}{2}-\inf_{(x_1,x_2)}\left( R(x_1,x_2)-\frac{\beta x_1^2}{2} \right)
\end{equation}

Physically, the value of $x_1$ and $x_2$ for a given $\Delta$ and $\beta$ would be such that they minimise $I(x_1,x_2)$. Let $m$ and $q$ represent the minimum value of $x_1$ and $x_2$ respectively for a given $\Delta$ and $\beta$. Minimising $I(x_1,x_2)$ with respect to $x_1$ and $x_2$ gives the following two equations for $m$ and $q$:

\begin{equation}
\label{mpure}
\tanh (\beta m)=\frac{m}{q}
\end{equation}

\begin{equation}
\label{qpure}
-\beta \Delta+\log[\frac{\sqrt{q^2-m^2}}{2(1-q)}]=0
\end{equation}
 These two equations together give the phase diagram of the system. Note that $\Delta$ does 
not enter Eq. \ref{mpure} directly, but it enters via Eq. \ref{qpure}. In the regime of second 
order transition, we can expand Eqs. \ref{mpure} and \ref{qpure} for small $m$ to get 
the transition point. Expanding Eq. \ref{mpure} for small $m$ and keeping first nontrivial term, we get
\begin{equation}
m^2=\frac{3}{\beta^3} \left(\beta-\frac{1}{q} \right)
\end{equation}
This implies that the order parameter $m$ becomes non zero only when $q > 1/\beta$. 
Hence, taking $q=\frac{1}{\beta}(1+\epsilon)$, we get $m^2=3 \epsilon/\beta^2$ and substituting in Eq. \ref{qpure}, we get to leading order in $\epsilon$,
\begin{equation}
(1-4 e^{\beta \Delta}) \epsilon+2(1-2(\beta-1) e^{\beta \Delta})=0
\end{equation}

This gives the critical point of continuous transition to be

\begin{equation}
\label{cppure}
e^{\beta \Delta} =\frac{1}{2 (\beta-1)}
\end{equation}

This gives $\beta_c=3/2$ for $\Delta=0$, as expected. The linear approximation will break down when $e^{\beta \Delta}=1/4$. Substituting this in the above equation will give the tricritical point. We get $\beta_{tcp}=3$. For $\beta > \beta_{tcp}$, linear approximation breaks down and system will have first 
order transition. Value of crystal field at this point would be:
\begin{equation}
\Delta_{tcp}=-\frac{\log[4]}{3}
\end{equation}
For $\Delta<\Delta_{tcp}$ system will undergo first order transition and for $\Delta>\Delta_{tcp}$, there is a line of continuous transitions given by 
Eq. \ref{cppure}. As $\beta \rightarrow \infty$, there are two possibilities: $m=q=1$
or $m=q=0$(see Eq. 21) with energies $-(0.5+\Delta)$ and $0$
respectively. Hence for $\Delta \leq -0.5$, the disordered state always
wins. Hence the line of first order transitions ends at
$\Delta_c=-0.5$ as $\beta \rightarrow \infty$. For the rest of the first order line, we obtain 
the global minima of $I(x_1,x_2)$ for a given $(\beta,\Delta)$ numerically. The fixed point equations, Eq. \ref{mpure} and \ref{qpure} are exactly the same as obtained by Blume Emery and Griffiths \cite{beg}. Hence, we obtain the known phase diagram of the pure Blume-Capel model. Also taking $\Delta=0$ in Eq. \ref{entropy} gives the entropy of the system in the microcanonical ensemble \cite{barre}.

\section{Random Disorder Case}

Let us assume that the random crystal fields come from an i.i.d. bimodal distribution of the kind:

\begin{equation}
P(\Delta_i)= [p\delta(\Delta_i-\Delta)+(1-p) \delta(\Delta_i+\Delta)]
\end{equation}
For $p=0,1$ there is no disorder and the maximum effect of disorder will be at $p=1/2$. $p=1$ will be same as the pure case discussed in the previous section and $p=0$ will have  the same behaviour, but at $-\Delta$. For example, $p=0$ there is a first order transition at $\Delta=1/2$ as $\beta \rightarrow \infty$.

\subsection{Rate Function}

We again first calculate the rate function for the $H_{ni}=-\sum \Delta_i s_i$. Logarithmic moment generating function of $N Y_N=(N S_N,N \sigma_N)$ is given by Eq. 4 and $\lim\limits_{N\rightarrow\infty}\frac{1}{N}\Lambda_N(x_1,x_2)=f(x_1,x_2)-f(0,0)$, where $f(x_1,x_2)$ is now the expectation value of $f_N(x_1,x_2)$ as defined in Eq. 6, with respect to the distribution of the random fields. This can be achieved by using strong law of large numbers on Eq. 6 to replace the sum by the expectation value. For bimodal random crystal fields we get,
\begin{equation}
f(x_1,x_2) = p \log[2 e^{x_2+\beta \Delta} \cosh x_1 +1]+(1-p) \log[2 e^{x_2-\beta \Delta} \cosh x_1+1]
\end{equation}
The rate function $R(x_1,x_2)$ will be given by Eq. 7. Let us again define, $W(y_1,y_2)=x_1 y_1+x_2 y_2-\Lambda(y_1,y_2)$. We get,

\begin{eqnarray}
W(y_1,y_2) &=& x_1y_1+x_2y_2-p \log[1+e^{y_1+y_2+\beta \Delta}+e^{-y_1+y_2+\beta \Delta}]\\\nonumber &-& (1-p) \log[1+e^{y_1+y_2-\beta \Delta}+e^{-y_1+y_2-\beta \Delta}] \\ \nonumber &+& p\log[1+2 e^{\beta \Delta}]+(1-p) \log[1+2 e^{-\beta \Delta}]
\end{eqnarray}

Minimising $W(y_1,y_2)$ with respect to $y_1$ and $y_2$ and substituting back the 
resulting $y^*_1$ and $y^*_2$ in $W(y_1,y_2)$ will give the rate function $R(x_1,x_2)$. 
Equations for $y^*_1$ and $y^*_2$ are:
\begin{eqnarray}
x_1&=& 2 e^{y^*_2} \sinh y^*_1 \left(\frac{p e^{\beta \Delta}}{1+2 e^{y^*_2+\beta \Delta} \cosh y^*_1}+\frac{(1-p) e^{-\beta \Delta}}{1+2 e^{y^*_2-\beta \Delta} \cosh y^*_1}\right) \\
x_2 &= & 2 e^{y^*_2} \cosh y^*_1 \left(\frac{p e^{\beta \Delta}}{1+2 e^{y^*_2+\beta \Delta} \cosh y^*_1}+\frac{(1-p) e^{-\beta \Delta}}{1+2 e^{y^*_2-\beta \Delta} \cosh y^*_1}\right)
\end{eqnarray}
It is easier to work with a new variable $z= 2 e^{y^*_2} \cosh(y^*_1)$ rather than 
$y^*_2$ directly. On solving the above two equations and substituting, we get the rate function $R(x_1,x_2)$ to be

\begin{eqnarray}
R(x_1,x_2)&=&x_1 \tanh^{-1} \frac{x_1}{x_2}+x_2[\log(z)-\log(2\cosh(\tanh^{-1} (x_1/x_2))]]\\ \nonumber
& -& p \log(1+z e^{\beta \Delta})-(1-p) \log(1+z e^{\beta \Delta})+p\log(1+2 e^{\beta \Delta})\\\nonumber
&+& (1-p) \log(1+2 e^{\beta \Delta})
\end{eqnarray}
where $z$ is a function independent of $x_1$ but depends only on $x_2,p,\beta$ and $\Delta$, is a solution to the 
quadratic equation
\begin{equation}
\label{zeq}
\frac{x_2}{z}=\frac{p e^{\beta \Delta}}{1+z e^{\beta \Delta}}+\frac{(1-p) e^{-\beta \Delta}}{1+z e^{-\beta \Delta}}
\end{equation}

Hence the rate function $I(x_1,x_2)$ of the full Hamiltonian would again be given by Eq. 11. The minimum will give the value of $m$ and $q$ for a given $p,\beta$ and $\Delta$. Minimising $I(x_1,x_2)$, we get the following two equations for $m$ and $q$:

\begin{equation}
\label{opeq1}
\tanh(\beta m)=\frac{m}{q}
\end{equation}
and
\begin{equation}
\label{opeq2}
z=\frac{2}{\sqrt{1-m^2/q^2}}
\end{equation}
where $z$ satisfies the following equation:
\begin{equation}
\label{zeq}
\frac{q}{z}=\frac{p e^{\beta \Delta}}{1+z e^{\beta \Delta}}+\frac{(1-p) e^{-\beta \Delta}}{1+z e^{-\beta \Delta}}
\end{equation}

Interestingly, again $\Delta$ does not enter Eq. \ref{opeq1}, which is the equation governing the behaviour of magnetisation. Again, if there is a second order transition, then that can be evaluated by expanding Eq. \ref{opeq1} and \ref{opeq2} for small $m$. Expanding Eq. \ref{opeq1} for small $m$, we get
\begin{equation}
m^2=\frac{3}{\beta^3} \left(\beta-\frac{1}{q} \right)
\end{equation}
This implies that the order parameter $m$ becomes non zero only when $q > 1/\beta$. Hence, if 
there is a continuous transition, it will occur at $q=1/\beta$ and $m=0$. Putting these two 
values in Eq. \ref{opeq2} gives $z_c=2$ at the transition point. We can then calculate the point of 
continuous transition by evaluating Eq. \ref{zeq}, with $z=2$ and $q=1/\beta$. This gives:

\begin{equation}
\label{cpd}
5-4 \beta = 2(\beta p -1) e^{\beta \Delta}+2 (\beta -\beta p-1) e^{-\beta \Delta} 
\end{equation}
Note that for $p=1/2$ the equation is insensitive to the sign of $\Delta$. As expected, $p=0$ and $p=1$ correspond to the pure case. Transition point for a given $\Delta$ for $p=0$ is the same as that for $-\Delta$ with $p=1$. There is a symmetry about $p=1/2$, hence it is better to work with $\delta$ such that $p=1/2-\delta$, ($-1/2 \le \delta \le 1/2$).  Hence, $\delta =0$ corresponds to the maximum disorder and as we move away from $\delta=0$,  there is a asymmetry in the distribution . Hence, the transition is same for $(\delta,\Delta)$ and $(-\delta,-\Delta)$.  We will hence study $0\le \delta \le 1/2$ and the phase diagram for $-1/2\le \delta \le 0$ can be obtained by reversing the sign of $\Delta$.

We get the equation for the point of continuous transition to be:
\begin{equation}
\label{cpeq}
5-4 \beta =2 (\beta-2) \cosh(\beta \Delta)-4 \beta \delta \sinh(\beta \Delta)
\end{equation}

Expanding around the critical point by taking $q=(1+\epsilon)/\beta$, we get

\begin{equation}
\epsilon=\frac{5-4 \beta -2 (\beta-2) \cosh(\beta \Delta)+4 \beta \delta \sinh(\beta \Delta) }{-17+12 \beta+(3\beta-10) \cosh(\beta \Delta)-6 \beta \delta \sinh(\beta \Delta)}
\end{equation}

The linear approximation will break down when
\begin{equation}
\label{linearb}
-17+12 \beta+(3\beta-10) \cosh(\beta \Delta)-6 \beta \delta \sinh(\beta \Delta)=0
\end{equation}

Simultaneous solution of Eq. \ref{cpeq} and \ref{linearb} will hence give the tricritical point. 
Beyond this point either the system will have first order transition or no transitions at all. 
Solving them together we get:

\begin{equation}
\label{tcp}
\cosh(\beta \Delta)=\frac{12 \beta -19}{8}
\end{equation}
Note that since $\cosh(\beta \Delta) \ge 1$ for all real values of $\beta$ and $\Delta$, the tricritical point 
can occur only if $\beta_c$ given by Eq. \ref{cpeq} is greater than or equal to $9/4$ for some $\Delta$.

\subsection{Phase Diagram}

For $\Delta=0$, Eq. \ref{cpeq} gives $\beta_c=3/2$ independent of $\delta$. Also for $\delta=1/2$ and $\delta=-1/2$ one recovers the phase diagram of the pure case. For example, for $\delta=1/2$, solving Eq. \ref{cpeq} and \ref{linearb}, one gets the 
tricritical point to be at $\beta_c=3$, which gives $\Delta_{tcp}$ to be:

\begin{equation}
\Delta_{tcp} =\frac{\cosh^{-1}17/8}{3}=\log(4)/3
\end{equation} 
Which as expected, is equal to $-\Delta_{tcp}$ of the pure case discussed in Section 3. For $-\infty \le \Delta <\Delta_{tcp}$, there is a line of critical points given by Eq. \ref{cpeq}. Beyond $\Delta_{tcp}$, we have to look for the global minima of $I(x_1,x_2)$. As expected, for $\Delta>0.5$ , the minima gets shifted to $m=0$ for all values of $\beta$. The resultant phase diagram is plotted in Fig 1(d).

The maximum effect of disorder will be at $\delta=0$. Substituting $\delta=0$ in Eq. \ref{cpeq}, we get a line of continuous transitions, given by:
\begin{equation}
5-4 \beta = 2 (\beta -2) \cosh(\beta \Delta)
\end{equation}

As expected, this equation is insensitive to the sign of $\Delta$. Substituting $\cosh \beta \Delta = (12 \beta -19)/8$, in the above equation gives only imaginary values of $\beta$, implying that the above equation is valid for all values of $\Delta$ and $\beta$. Hence, the system has a continuous transition for all values of $\Delta$, with transition point increasing from $\beta_c=3/2$ to $\beta_c=2$ as one increases $|\Delta|$ from 0 to $\infty$(See Fig 1(a)). This can be verified by looking at the full rate function $I(x_1,x_2)$ we well.

So we see that at $\delta=0$ the system exhibits a tricritical point, but for $\delta=1/2$, the transition gets rounded  off to a second order transition. What happens for in between values? For that on substituting 
Eq. \ref{tcp} into Eq. \ref{cpeq}, we get a quartic equation in $\beta$, whose solution would give 
the tricritical point for a given $\delta$. On numerically solving the resulting equation, we find that for $\delta<\delta_{th} =0.4547725\pm0.0000001$, there is no real $\beta$ as a  solution, implying that the system has only continuous transition for $|\delta|<|\delta_{th}|$. At $\delta=0.4547725$, system has a tricritical point at $\Delta_{tcp} = 0.474$ with $\beta_{tcp} =4.68$. For $\delta_{th} <\delta \le 1/2$, the random system has a tricritical point and the phase diagram looks similar to the pure case. In Fig 1(c) and 1(d) we have plotted the phase diagram for $\delta=0.48(p=0.02)$ and $\delta=1/2(p=0)$ for comparison. The line of first order transition is obtained by looking at global minima of the full rate function numerically, at the fixed points given by  Eqs. \ref{opeq1}, \ref{opeq2} and \ref{zeq}. 

The line of first order transition ends at zero temperature at a finite value of $\Delta$ for $\delta_{th} < \delta \le 1/2$.  We find that there is no ordered phase for $\Delta> 1-\delta$, for $\delta>\delta_{th}$,  even at zero temperature. But there is a continuous transition for all values of  $\Delta <\Delta_{tcp}$, i.e  the line of continuous transition extends to $\Delta =-\infty$. As $\Delta \rightarrow -\infty$, for all values $0 \le \delta \le 1/2$, the point of continuous transition $\beta_c \rightarrow 2/(1+2 \delta)$. For $0<\delta <\delta_{th}$, there is a continuous transition for all positive values of $\Delta$ as well with $\beta_c =2/(1-2 \delta)$ as $\Delta \rightarrow \infty$(see Fig. 1(b)).

\begin{figure}
\centering
\begin{tabular}{cc}
\includegraphics[width=4.4cm]{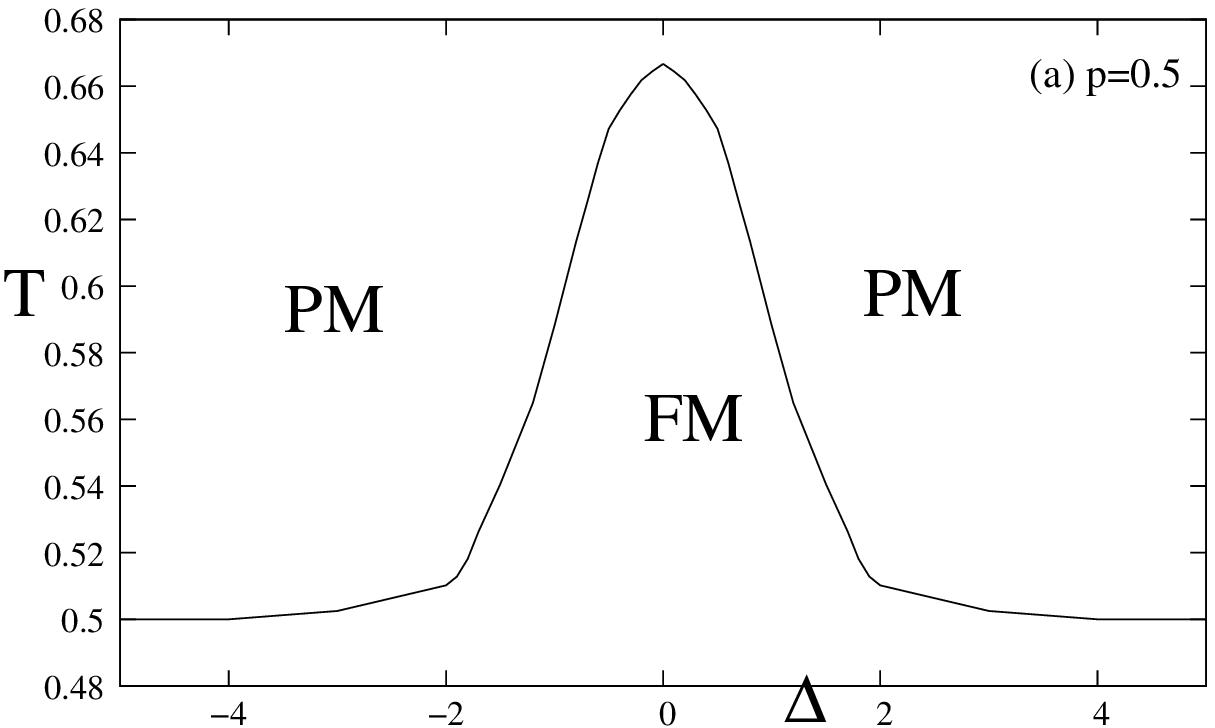} &
\includegraphics[width=4.4cm]{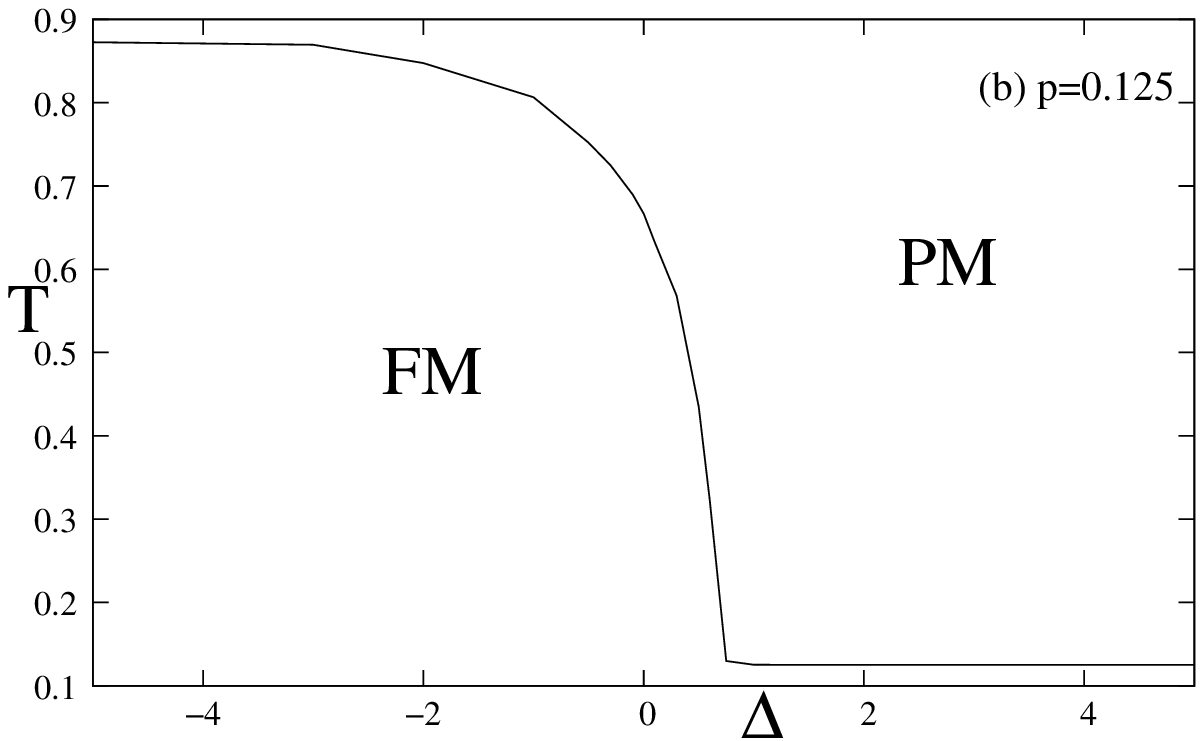}\\ 
\includegraphics[width=4.4cm]{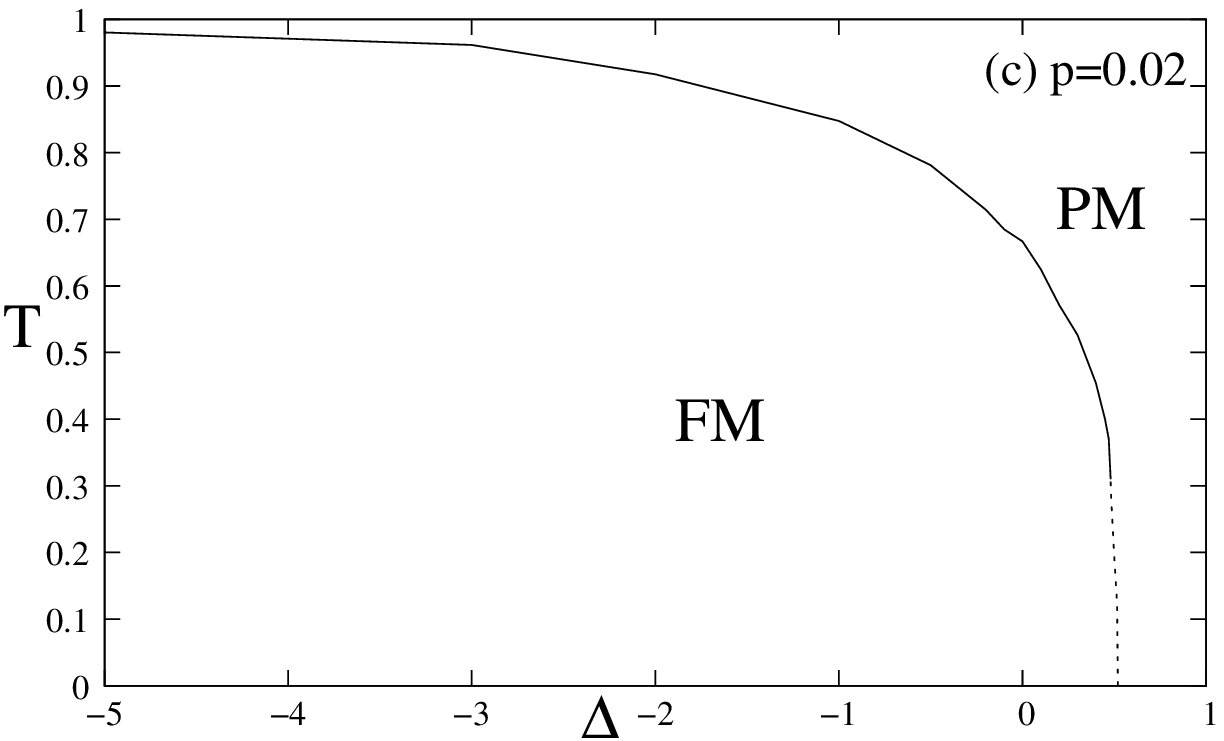} &
\includegraphics[width=4.4cm]{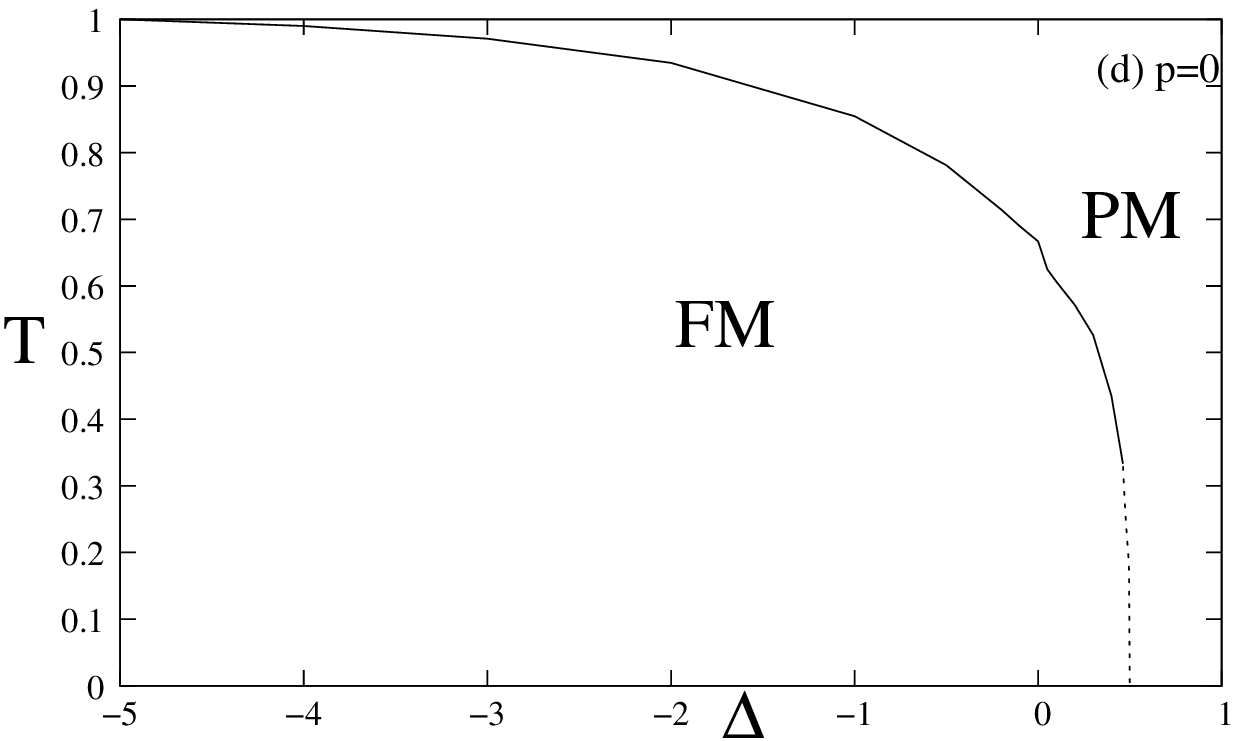}\\ 
\end{tabular}
\caption{Phase diagram of the model for few representative values of $p$ in ($T=1/\beta$,$\Delta$) plane. Solid line represents a line of continuous transitions from ferromagnetic(FM) 
to paramagnetic(PM) phase and the dotted line(in (c) and (d)) represents first order transitions. Fig 1(a) is for the maximum disorder case, $p=1/2(\delta=0)$ and 1(b) is for  $p=0.125(\delta=0.375)$, which is less than $\delta_{th}\approx 0.454$.  Fig 1(c) is for weak disorder, $p=0.02(\delta=0.48)$ and 1(d) is the phase diagram for $p=0(\delta=1/2)$, which corresponds to the pure case.}
\label{ratef}
\end{figure}

\section{Discussion}

Random crystal field Blume-Capel model has been studied using many approximate methods, but the results from different approximations do not match especially at low temperatures. For example, the solution obtained using pair approximation \cite{lara} does not have the symmetry around p. Similarly, other approximations \cite{mfdilution,santos}, predict a finite transition temperature for all non zero strengths of disorder, which has been proven incorrect by earlier studies using effective field theory \cite{kaneyoshi}. This is because, these treatments deal with only one order parameter ($m$) and the entropy of zero spin state is not fully accounted for in these approximations. In contrast, the free energy(given by minima of the rate function) from our calculations, takes care of the entropic contribution correctly. And we find that for weak disorder, just like the pure case there is no ordered phase, beyond a certain value of $\Delta$. For  $p=1/2$, our results match with the prediction using effective field theories\cite{yuksel}.  We find that unlike in two dimensions\cite{branco}, where the transition is a continuous transition for all values of $\Delta$ and $p$, on a fully connected graph this happens only for strong enough disorder( $0.046< p <0.954$). 

 In general, studying the model numerically even in the pure case for three dimensions is challenging \cite{zierenberg}, and the introduction of random fields would make it even harder as, in general it is hard to study quenched disorder via simulations due to lack of self averaging \cite{fytas}. Hence, solution on fully connected graph gives us an understanding of the effect of disorder for $d>2$. 

In case of $p$-spin model, it was shown \cite{sumedha} that the rate function is similar to Landau free energy functional. Expanding $I(x_1,x_2)$ for the pure case gives the known form of Landau functional for Blume-Capel model \cite{huang}. The method presented in this paper is very general and can be easily extended to higher spins \cite{spin2}. In a related model, namely the Blume-Emery-Griffiths model, recently it was shown that though the transition becomes a continuous transition, it does not fall in the Ising universality class\cite{berker1}. It would be interesting to see if the same holds on a complete graph. There are other models as well like Potts model \cite{fernandez}, 3-color model \cite{bellafard,zhu} and 4-color \cite{ashkin} Ashkin-Teller model, which have been studied numerically in the presence of quenched disorder in three dimensions. It would be straightforward to get the behaviour of these models on a complete graph using our approach. It should also be possible to extend the method to study the models with continuous spins as well. 

\section*{Appendix}

The tilted large deviation technique \cite{hollander} gives LDP for a new sequence of probability measures generated by another sequence of probabilities satisfying an LDP. It says the following:

\begin{thm}[Tilted LDP]
	Let $\{P_n\}_n$ be a sequence of probability measures on a complete separable metric space $\mathcal{X}$ satisfying LDP with rate $n$ and rate function $I:\mathcal{X}\rightarrow\mathbb{R}$. Let $F:\mathcal{X}\rightarrow \mathbb{R}$ be a continuous function bounded from above. For Borel subset $A$ of $\mathcal{X}$, define $J_n(A)=\int_{A} e^{nF(x)}P_n(dx)$ and $P_n^F(A)= \frac{J_n(A)}{J_n(\mathcal{X})}$. Then the sequence of probability measures $\{P_n^F\}_n$  on $\mathcal{X}$ satisfies LDP with rate $n$ and rate function $$I^F(x)=\sup_{y\in \mathcal{X}}[F(y)-I(x)]-[F(x)-I(x)].$$
\end{thm}

\noindent{\em Proof:} Let $C$ be a closed subset of $\mathcal{X}$. Then 
\begin{equation}
\limsup\limits_{n\rightarrow\infty}\frac{1}{n}\log P_n^F(C)=\limsup\limits_{n\rightarrow\infty}\frac{1}{n}\log J_n(C)-\sup\limits_{x\in\mathcal{X}}[F(x)-I(x)],
\end{equation}
 since by Varadhan's Lemma \cite{varadhan},
\begin{equation}
\lim\limits_{n\rightarrow\infty}\frac{1}{n}\log J_n (\mathcal{X})=\sup\limits_{x\in\mathcal{X}}[F(x)-I(x)].
\end{equation}

Let $a(C)=\sup\limits_{x\in C}F(x)$ and $b(C)=\sup\limits_{x\in C}[F(x)-I(x)]$. Then note that $-\infty<b(C)\leq a(C)<\infty$.

Moreover, let $$D(C)=F^{-1}([b(C),a(C)])\cap C$$ and for $N\in\mathbb{N}$, $$D_j^N(C)=F^{-1}([\alpha_{j-1}^N,\alpha_j^N])\cap C, \quad j=1,\cdots,N,$$
where $\alpha_j^N=b(C)+\frac{j}{N}(a(C)-b(C))$.

Since, $D_j^N(C)$ are closed sets, by definition of LDP, $$\limsup\limits_{n\rightarrow\infty}\frac{1}{n}\log P_n(D_j^N(C))\leq -\inf\limits_{x\in D_j^N(C)}I(x), \quad \forall j.$$

Hence $$\limsup\limits_{n\rightarrow\infty}\frac{1}{n}\log J_n(D(C))\leq \max\limits_{1\leq j\leq N}\limsup\limits_{n\rightarrow\infty}\frac{1}{n}\log J_n(D_j^N(C)).$$

Since $F(x)\leq \alpha_j^N$ on $D_j^N(C)$ and $\alpha_j^N \leq \inf\limits_{x\in D_j^N(C)}F(x)+\frac{1}{N}(a(C)-b(C)) \leq \sup\limits_{x\in D_j^N(C)}F(x)+\frac{1}{N}(a(C)-b(C))$, 
$$\limsup\limits_{n\rightarrow\infty}\frac{1}{n}\log J_n(D_j^N(C))\leq \sup\limits_{x\in D_j^N(C)}F(x)-\inf\limits_{x\in D_j^N(C)}I(x)+\frac{1}{N}(a(C)-b(C))$$
and hence
$$\limsup\limits_{n\rightarrow\infty}\frac{1}{n}\log J_n(D(C))\leq \sup\limits_{x\in D(C)}[F(x)-I(x)]+\frac{1}{N}(a(C)-b(C)).$$
Thus $N$ being arbitrary, we get,
$$\limsup\limits_{n\rightarrow\infty}\frac{1}{n}\log J_n(D(C))\leq b(C).$$
For the set $C\setminus D(C)$, we have $J_n(C\setminus D(C))\leq e^{nb(C)}$. Hence we have,
$$\limsup\limits_{n\rightarrow\infty}\frac{1}{n}\log J_n(C)\leq b(C).$$

Now let $O$ be an open subset of $\mathcal{X}$. Then $$\liminf\limits_{n\rightarrow\infty}\frac{1}{n}\log P_n^F(O)=\liminf\limits_{n\rightarrow\infty}\frac{1}{n}\log J_n(O)-\sup\limits_{x\in\mathcal{X}}[F(x)-I(x)].$$ 
For arbitrary $\epsilon>0$ and $x\in O$, set $$O_{x,\epsilon}:=\{y\in O:F(y)>F(x)-\epsilon\}.$$ Note that $O_{x,\epsilon}$ is an open set containing $x$ and by LDP 
$$\liminf\limits_{n\rightarrow\infty}\frac{1}{n}\log P_n(O_{x,\epsilon})\geq -\inf\limits_{x\in O_{x,\epsilon}}I(x)\geq  -I(x).$$
Hence 
$$\liminf\limits_{n\rightarrow\infty}\frac{1}{n}\log J_n(O)\geq F(x)-\epsilon-I(x).$$
$\epsilon>0$ being arbitrary and taking supremum over all $x\in O$, we get,
$$\liminf\limits_{n\rightarrow\infty}\frac{1}{n}\log J_n(O)\geq \sup\limits_{x\in O}[F(x)-I(x)].$$
Hence the proof $\blacksquare$.

Now in the Blume-Capel context, let us consider the probability of the configuration $C_N$ when the full Hamiltonian (Eq. 1) is considered be
\begin{equation}
P_{N,\beta}(C_N)\propto \exp\{-\beta H_N(C_N)\}.
\end{equation}

Then note that $P_{N,\beta}\circ Y_N^{-1}$ is a tilted version of $Q\circ Y_N^{-1}$. To see this, let us consider the bounded continuous function $F:\mathbb{R}^2 \rightarrow \mathbb{R}$ as defined in Eq. 10 and define $\mathcal{X}_N=\cup_{k=0}^N \mathcal{X}_N^k$ where $\mathcal{X}_N^k=\{-\frac{k}{N},-\frac{k-2}{N},\cdots,\frac{k-2}{N},\frac{k}{N}\} \times \{\frac{k}{N}\}$. Moreover if we define $1_A:A\rightarrow \{0,1\}$ as $$1_A(x)=\begin{cases}
1 & \mbox{if } x\in A,\\0 &\mbox{if } x\notin A,\end{cases}$$ then for any Borel set $A$ of $\mathbb{R}^2$,
$$\begin{array}{rcl}
	P_{N,\beta}\circ Y_N^{-1}(A)&=&\sum\limits_{k=0}^N\sum\limits_{m=-k}^{k}1_{A\cap\mathcal{X}_N^k}(\frac{m}{N},\frac{k}{N}) P_{N,\beta}(\{NY_N=(m,k)\})\\
	&=&\sum\limits_{k=0}^N\sum\limits_{m=-k}^{k}1_{A\cap\mathcal{X}_N^k}(\frac{m}{N},\frac{k}{N})\sum\limits_{S_N:NY_N(S_N)=(m,k)} P_{N,\beta}(S_N)\\
	&=&\frac{\sum\limits_{k=0}^N\sum\limits_{m=-k}^{k}1_{A\cap\mathcal{X}_N^k}(\frac{m}{N},\frac{k}{N})e^{NF(\frac{m}{N},\frac{k}{N})}\sum\limits_{S_N:NY_N(S_N)=(m,k)} \prod\limits_{i=1}^{N}Q_i(s_i)}{\sum\limits_{k=0}^N\sum\limits_{m=-k}^{k}1_{\mathcal{X}_N^k}(\frac{m}{N},\frac{k}{N})e^{NF(\frac{m}{N},\frac{k}{N})}\sum\limits_{S_N:NY_N(S_N)=(m,k)} \prod\limits_{i=1}^{N}Q_i(s_i)}\\
	&=& \frac{\sum\limits_{k=0}^N\sum\limits_{m=-k}^{k}1_{A\cap\mathcal{X}_N^k}(\frac{m}{N},\frac{k}{N})e^{NF(\frac{m}{N},\frac{k}{N})}Q\circ Y_N^{-1}(\frac{m}{N},\frac{k}{N})}{\sum\limits_{k=0}^N\sum\limits_{m=-k}^{k}1_{\mathcal{X}_N^k}(\frac{m}{N},\frac{k}{N})e^{NF(\frac{m}{N},\frac{k}{N})}Q\circ Y_N^{-1}(\frac{m}{N},\frac{k}{N})},
\end{array}$$

Hence by tilted LDP, we get the following result:
\begin{thm}
	Almost surely, the sequence $\{Y_N\}_N$ satisfies LDP w.r.t. $P_{N,\beta}$ with rate $N$ and rate function 
	\begin{equation}
	I(x_1,x_2)=\begin{cases}R(x_1,x_2)-\frac{1}{2}\beta x_1^2-\inf\limits_{(y_1,y_2)\in \mathbb{R}^2}\{R(y_1,y_2)-\frac{1}{2}\beta y_1^2\},& |x_1| \leq x_2 \leq 1,\\
	\infty, &\mbox{ otherwise}
	\end{cases}
	\end{equation}
\end{thm}
 The above technique is similar to that of L\"{o}we et al\cite{lowe}.


\begin{thebibliography}{99}
\bibitem{harris} A.B. Harris, J. Phys. C, {\bf{7}},1671(1974).
\bibitem{imrywortis} Y. Imry and M Wortis, Phys. Rev. B, {\bf 19} 3580,(1978)
\bibitem{aizenman} M Aizenman and J Wehr,Phys. Rev. Let, {\bf 62} 2503, (1989)
\bibitem{huiberker} K. Hui and A.N. Berker, Phys. Rev. Lett. {\bf 62},2507, (1989)
\bibitem{branco} N.S. Branco and B.M. Boechat, Phys. Rev. B {\bf 56},11673(1997)
\bibitem{greenblatt} R. L. Greenblatt, M. Aizenman and J. L. Lebowitz, Phys. Rev. Lett. {\bf 103} 
197201(2009).
\bibitem{cardy}J Cardy, Physica A {\bf 263} 215,1999; J. Cardy and J. L. Jacobsen, Phys. Rev. Lett. {\bf 79}, 4063(1997).
\bibitem{blume} M Blume, Phys. Rev. {\bf 141} 517(1966)
\bibitem{capel} H. W. Capel, Physica(Utrecht) {\bf 33},295(1967).
\bibitem{beg} M . Blume, V.J. Emery and R. B. Griffiths, Phys. Rev. A {\bf 4} 1071(1971).
\bibitem{frazer} B. C. Frazer,  G. Shirane, D. E. Cox and C. E. Olsen, J. Appl. Phys. {\bf 37} 1386(1966).
\bibitem{helium} C. Buzano, M. Cieplak, M. R. Swift, F. Toigo and J. R. Banavar, Phys. Rev. Lett. {\bf 69} 221(1992)
\bibitem{buzano} C. Buzano, A Maritan, A. Pelizzola, J. Phys. Cond. Matt. {\bf 6} 327(1994)
\bibitem{schupper} N Schupper and N. M. Shnerb, Phys. Rev. E {\bf 72}, 046107(2005).
\bibitem{bulbul} S. Sarkar and B. Chakraborty, Phys. Rev. E {\bf 91}, 042201(2015).
\bibitem{mfdilution} A Benyoussef,T Biaz,M Saber and M Touzani J. Phys. C {\bf 20} 5349(1987);E. Albayrak  Physica A{\bf 392} 552(2013).
\bibitem{yuksel} Y Yuksel, U Akinci and H Polat, Physica A {\bf 391} 2819(2012).
\bibitem{bethe} E. Albayrak, Physica A {\bf 390} 1529(2011)
\bibitem{lara} D. P. Lara, Revista Mexicana de Fiscia {\bf 58} 203(2012).

\bibitem{snowman} D. P. Snowman, Phys. Rev. E {\bf 79} 041126(2009).
\bibitem{santos} P.V. Santos, F.A de Costa and J.M. de Araujo, Physics Letters A{\bf 379} 1397(2015).
\bibitem{kaneyoshi} T. Kaneyoshi and J. Mieinick, J. Phys.: Con. Matt. {\bf 2} 8773(1990).
\bibitem{zierenberg} J Zierenberg, N. G. Fytas and W. Janke, Phys. Rev. E {\bf 91} 032126(2015)
\bibitem{fernandez} L.A. Fernandez, A G Guerrero, V Martin-Mayor and J.J. Ruiz Lorenzo, Phys. Rev. Lett, {\bf 100} 057201(2008)
\bibitem{fytas} N. G. Fytas and A. Malakis, EPJ B {\bf 79}, 13(2011); A. Malakis and N. G. Fytas, Phys. Rev. E {\bf 73}, 016109(2006).
\bibitem{touchette} H. Touchette, Physics Reports, {\bf 478}, 1(2009).

\bibitem{ellis} R.S. Ellis, Peter T Otto and H. Touchette, Annals of Applied Probability,{\bf 15},2203-2254(2005).
\bibitem{barre} J. Barre, D. Mukamel and S. Ruffo, Phys. Rev. Lett. {\bf 87} 030601, 2001.
\bibitem{sumedha} Sumedha, and Sushant K. Singh, Physica A {\bf 442} 276(2016).
\bibitem{lowe} M Lowe, R Meiners and F Torres, J. Phys. A:Math,{\bf 46}, 125004 (2013).
\bibitem{hollander} Theorem III.17 of Frank den Hollander, Large Deviations, Fields Institute Monographs,AMS(2000)
\bibitem{huang} Look for example in "Introduction to Statistical mechanics" by K. Huang,
Wiley; 2 edition (May 13, 1987)
\bibitem{spin2} L. Bahmad, A. Benyoussef and A. El Kenz, Phys. Rev. B {\bf 76} 094412(2007).

\bibitem{berker1} A. Falicov and A.N. Berker, Phys. Rev. Lett. {\bf 76}, 4380(1996);A. Malakis, A. Nihat Berker, N.G. Fytas and T. Papakonstantinou, Phys. Rev. E{\bf 85} 061106(2012).





\bibitem{bellafard} A. Bellafard, S. Chakravarty, M. Troyer, H.G. Katzgraber, Annals of Phys. {\bf 357}, 66(2015).
\bibitem{zhu} Q. Zhu, X. Wan, R. Naryanan, J.A. Hoyos, T. Vojta, Phys. Rev. B {\bf 91}, 224201(2015)
\bibitem{ashkin}A. K. Imrahim and T. Vojta, arXiv:1602.04508
\bibitem{varadhan} Theorem III.13 of Frank den Hollander, Large Deviations, Fields Institute Monographs,AMS(2000)

\end{thebibliography}
\end{document}